\documentclass[twocolumn,floatfix, showpacs]{revtex4}

\usepackage[final]{epsfig}
\usepackage{amsmath}
\begin{document}
 
\title{On the sensitivity of jet quenching to near $T_C$ enhancement of the medium opacity}
 
\author{Thorsten Renk}
\email{thorsten.i.renk@jyu.fi}
\affiliation{Department of Physics, P.O. Box 35, FI-40014 University of Jyv\"askyl\"a, Finland}
\affiliation{Helsinki Institute of Physics, P.O. Box 64, FI-00014 University of Helsinki, Finland}


\begin{abstract}
One of the main goals of the study of high transverse momentum ($P_T$) observables in the context of ultrarelativisic heavy-ion collisions is the determination of properties of the produced QCD matter. In particular, the transport coefficients $\hat{q}$ and $\hat{e}$, characterizing the interaction of the medium with a high $p_T$ parton, are  accessible via high $P_T$ probes. However, a precision extraction of their temperature dependence from current data faces the problem that neither the spacetime geometry of the evolving matter droplet nor the link between thermodynamics and transport coefficients is unambiguously known, and various conjectured scenarios how thermodynamics and transport coefficients behave close to the phase transition exist. While often a behaviour with the energy density $\hat{q} \sim \epsilon^{3/4}$ is assumed which leads to a parametric decrease of the scaled $\hat{q}(T)/T^3$ close to the critical temperature $T_C$, other scenarios expect insted a near $T_C$ enhancement of jet quenching. In this work, the systematic response of both the extraction of $\hat{q}$ and $v_2$ at high $P_T$ to modification of jet quenching close to $T_C$ is systematically investigated within YaJEM, a well-tested in-medium shower evolution Monte-Carlo (MC) code. 
\end{abstract}
 
\maketitle

\section{Introduction}

High $P_T$ observables are a cornerstone of the experimental ultrarelativistic heavy-ion (A-A) program at both RHIC and LHC. One crucial goal of this program is the extraction of properties of the produced Quantum Chromodynamics (QCD) matter droplet, for instance in terms of characteristic transport coefficients and their temperature dependence. Two such coefficients, $\hat{q}$ (the mean gain in transverse momentum squared of a high $p_T$ parton per unit pathlength) and $\hat{e}$ (the mean longitudinal momentum loss of a high $p_T$ parton per unit pathlegth) are particularly relevant in this context. Here, $\hat{q}$ is responsible for energy loss from hard partons into medium-induced soft gluon radiation whereas $\hat{e}$ causes energy loss into non-perturbative medium degrees of freedom (see e.g. \cite{Review} for a review).

The only known way of extracting the temperature dependence of these coefficients from the data is to use perturbative QCD (pQCD) to compute the primary hard parton productiona and  then to embed a model for the parton-medium final state interaction into a fluid-dynamical simulation of the matter to obtain the medium-modification to the final state of the hard process. The transport coefficient is in this approach computed as a function of thermodynamical parameters in the fluid dynamics, e.g. energy density $\epsilon$, temperature $T$ or entropy density $s$ as well as potentially the flow vector $u^\mu$ of the medium relative to the c.m. frame of the collision. 

Based on the notion of counting the number density of potential scattering centers in an ideal gas, many jet quenching models assume

\begin{equation}
\label{E-1}
\hat{q} \sim T^3  \text{,}\quad \hat{q} \sim s \quad \text{or} \quad \hat{q} \sim \epsilon^{3/4}
\end{equation}  which for an ideal gas equation of state $\epsilon = 3p$ all coincide. Differences to the ideal gas for all three expressions occur close to the phase transition and in the hadronic phase and have previously been discussed in e.g. \cite{SysEloss}. In the presence of a finite flow value, a relativistic correction term dependent on the local flow rapidty $\rho$ at the position $\zeta$ and the angle $\alpha$ between parton momentum vector and flow vector has been found both in weak and strong coupling as \cite{Flow1,Flow2}

\begin{equation}
\label{E-flow}
F(\rho(\zeta), \alpha(\zeta)) = \cosh \rho(\zeta) - \sinh \rho(\zeta) \cos\alpha(\zeta).
\end{equation}

In practice, this factor corresponds to a small correction \cite{FlowTest}. Given such a model setting, a $\chi^2$ fit of the proportionality constant between transport coefficient and thermodynamical parameter to e.g. the single inclusive hadron nuclear suppression factor $R_{AA}$ is possible \cite{PHENIX-fit} and results in a temperature dependence of $\hat{q}$ compatible with the data. However, such a procedure does not yield consistent results across different models \cite{Brick}, indicating that the uncertainties in the choice of the parton-medium interaction model and the medium evolution itself are substantial. One solution is to include other observables into the fit, for instance the suppression factor of the back-to-back coincidences $I_{AA}$ \cite{Armesto-fit}, however for multiple observables and across the full parameter space of available fluid dynamics and parton-medium interaction models, such a strategy becomes prohibitively expensive.

A further source of uncertainty is that any temperature dependence of the extracted transport coefficients is essentially not determined in a data-driven way but assumed {\itshape a priori} using a relation like Eq.~(\ref{E-1}). However, in \cite{Liao,Liao1} a scenario was suggested in which parton-medium interaction is not reduced but parametrically enhanced close to the phase transition temperature $T_C$ ('near $T_C$ enhancement', referred to as NTC in the following). This suggestion was driven by the need to explain the experimentally observed large split between in-plane and out of plane particle emission at high $P_T$ \cite{PHENIX-RP} when the dependence of $R_{AA}$ with respect to the reaction plane angle $\phi$ is considered in non-central A-A collisions. A careful systematic study across many different model frameworks \cite{SysFluid} has demonstrated however that the magnitude of the split is influenced by many factors, among them the assumed pathlength dependence of energy loss, the initial eccentricity of the medium, viscous entropy production during the medium evolution and the total size of the spacetime volume in which hard partons interact with the medium. Taking all these systematic uncertainties into account, it is not clear whether there is a remaining tension with the data, however there is a trend across several models to underpredict the spread \cite{SpreadData,CUJet,YaJEM-D}.

The aim of this work is to quantify the potential effect of a NTC scenario on both the extraction of a transport coefficient $\hat{q}$ and on the spread between in-plane and out of plane $R_{AA}$.

\section{The observable}

The observable considered in this study is the single inclusive hadron suppression factor $R_{AA}$ which is defined as the yield of high $P_T$ hadrons from an A-A collision normalized to the yield in p-p collisions at the same energy corrected for the number of binary collisions, 

\begin{equation}
\label{E-RAA}
R_{AA}(p_T,y) = \frac{dN^h_{AA}/dp_Tdy }{T_{AA}({\bf b}) d\sigma^{pp}/dp_Tdy}.
\end{equation}

The default expectation is $R_{AA} < 1$ in medium since parton-medium interaction is expected to lead to a flow of high $p_T$ parton momentum into medium degrees of freedom, thus effectively suppressing the yield in any given momentum bin. Nuclear initial state effects can however cause $R_{AA} > 1$ in some kinematical regions \cite{KinematicLimit}.

Experimentally, $R_{AA}$ can readily be obtained with respect to the angle $\phi$ of a hard hadron with the bulk matter $v_n$ event plane where $v_n$ is the $n$th coefficient in a harmonic expansion 
\begin{displaymath}
\frac{dN}{d\phi} = \frac{N}{2\pi} \left( 1 + \sum_n (2 v_n \cos(n\phi)) \right)
\end{displaymath}
 of the angular distribution of the bulk particle yield $dN/d\phi$. If $R_{AA}$ is obtained as a function of $\phi_2$, then the spread $S^{in}_{out}  = R_{AA}(0) - R_{AA}(\pi/2)$ between in plane and out of plane emission is an important observable sensitive to the medium geometry. 

Knowledge of $R_{AA}(0) = R_{AA}^{in}$  and $R_{AA}(\pi/2) = R_{AA}^{out}$ is approximately equivalent to angular averaged $R_{AA}$ and the second harmonic coefficient $v_2$ at high $P_T$, since \emph{if the modulation is a pure second harmonic} 
\begin{equation}
\label{E-v2}
R_{AA}^{in} = R_{AA} (1+2v_2) \quad \text{and} \quad R_{AA}^{out} = R_{AA} (1-2v_2)
\end{equation}
with $R_{AA}$ the angular averaged value.

Colloquially $v_2$ is frequently referred to as 'elliptic flow coefficient', but this is  highly misleading at high $P_T$ --- the angular modulation is not driven by any flow phenomenon but by the different strength of parton-medium interaction dependent on the density and size of traversed matter. The attenuation is known to be a non-linear function of the traversed length, in particular no matter how strong any interaction with the medium, $R_{AA} >0$ is always true.  This implies that for sufficiently low values of the average $R_{AA}$ and high $v_2$, Eq.~(\ref{E-v2}) can not be fulfilled and saturation leads to a distortion of the resulting angular structure from a pure $v_2$ modulation even if the matter distribution has a perfect second harmonic spatial eccentricity $\epsilon_2$. At the minimum of $R_{AA}$ at LHC at about $P_T = 10 $ GeV, this creates a spurious $v_4 \approx 0.2 \cdot v_2$ for the model used in this study.

Since this is commonly done in the literature, we will in the following discuss $\phi$  dependent physics nevertheless  in terms of $v_2$ with the above caveats in mind.

\section{Extraction of $\hat{q}$}

In the following, we parametrize NTC by an expression

\begin{equation}
\label{E-qhat}
\hat{q}(T) = 2 \cdot K \cdot T^3 \left[1 + c \cdot \exp\left(-\frac{(T-T_C)^2}{\sigma^2}\right) \right] F(\rho, \alpha)
\end{equation}

with $F(\rho,\alpha)$ as in Eq.~(\ref{E-flow}), $K$ a free parameter regulating the overall strength of the parton-medium interaction and $c, \sigma$ characterizing the strength and region of influence of the NTC. We test $\sigma = 10$ MeV and $\sigma = 30$ MeV in the following as well as $c$ in the range from 0 to 3.

Eq.~(\ref{E-qhat}) is applied to a 2+1d fluid dynamical simulation of the bulk matter evolution for 2.76 ATeV PbPb collisions \cite{SpectraLHC}. Using the local temperature $T$, the transport coefficient for every spacetime point dependent on the specific hard parton trajectory through the matter can be obtained. This medium is then probed by a hard partons in order to compute $R_{AA}$. We generate a distribution of hard partons based on leading order perturbative QCD expressions in a MC routine and initialize them in the transverse plane based on the binary collision probability distribution with a specified orientation with respect to the bulk $v_2$ event plane.

Parton-medium interaction is computed using the scenario YaJEM-DE \cite{YaJEM-DE} of the in-medium shower evolution code YaJEM \cite{YaJEM1,YaJEM2} which is an extension of the PYSHOW routine \cite{PYSHOW} simulating the QCD scale evolution in vacuum, and the reader interested in details of the simulation is referred to these works. YaJEM-DE is well tested against a number of different high $P_T$ observables both at RHIC and LHC kinematics, among them also observables with multiparticle final states, e.g. the dijet imbalance \cite{dijet1,dijet2}  and jet-hadron \cite{jet-h} correlations. However, it should be stressed that this is not of great relevance for the present work where we are interested in the relative change of qhat  and  $v_2$ due to NTC. Results of a systematic investigation of jet quenching with respect to different hydrodynamical background assumptions in \cite{SysFluid} indicate that the relative change in high $P_T$ $v_2$ driven by properties of the fluid dynamics is with an accuracy better than 10\% independent of the parton-medium interaction model used.

\begin{figure}
\epsfig{file=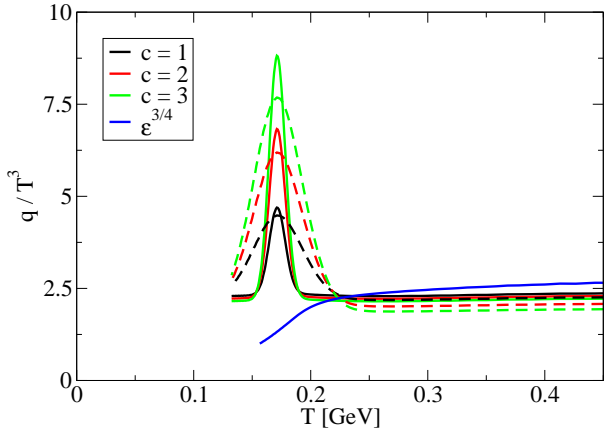, width=8cm}
\caption{\label{F-qhat}Temperature dependence of the scaled transport coefficient $\hat{q}/T^3$ for various near $T_C$ enhancement scenarios (see text) as deterimed by fits to $R_{AA}$ in 2.76 ATeV 0-10\% central Pb-Pb collisions at $P_T = 10$ GeV. Shown are $\sigma = 10 $ MeV (solid) and $\sigma = 30$ MeV (dashed). }
\end{figure}

$\hat{q}(T)$ is determined for any selection of $c$ and $\sigma$ by fitting $K$ in Eq.~(\ref{E-qhat}) to the angular averaged $R_{AA}$ in 0-10\% central collisions to ALICE charged hadron data \cite{ALICE-RAA} at $P_T = 10$ GeV . Fig.~\ref{F-qhat} shows the curves of $\hat{q}/T^3$ resulting from these fits for the various scenarios in comparison with the default ansatz $\hat{q} \sim \epsilon^{3/4} C(\rho, \alpha)$, assuming that jets decouple from bulk matter at the hypersurface characterized by $T_F = 0.13$ GeV.
From the results, it becomes clear that the high $T$ determination of $\hat{q}$ has no strong uncertainty associated with the near-$T_C$ behaviour of quenching. All scenarios find a stable value of $\hat{q}(T)/T^3 \sim 2.4$ to 2.5 GeV. This value is well in line with other model results \cite{CUJet,XNW}. Turning the argument around, one finds that as expected the near $T_C$ dynamics is not constrained by fitting angular averaged $R_{AA}$. 

The main uncertainty for a reliable determination of $\hat{q}$ still comes from the details of the fluid dynamical evolution \cite{Constraining}. For illlustration, assuming that hard partons decouple from the medium at a temperature of $T_F = 0.16$ GeV, the qualitative picture stays the same, but instead $\hat{q}(T)/T^3 \sim 3.8$ to 4.1 is found at high $T$ for the various scenarios. One can safely conclude that knowledge of matter geometry is currently the largest uncertainty for a determination of $\hat{q}$.

\section{Impact on $v_2$}

In order to assess the importance of NTC for the magnitude of $v_2$ at high $P_T$, we leave $\hat{q}(T)$ as determined by the mean $R_{AA}$ in central collisions as described above and use the same fluid dynamics computation for 30-40\% centrality. At $P_T = 10$ GeV, we compute $R_{AA}(\phi)$ and fit the expression
\begin{equation}
R_{AA}(\phi) = \langle R_{AA} \rangle \left(1 + 2 v_2 \cos(2 \phi)\right)
\end{equation}
to the result. As discussed above, this is not a perfect fit as there is a $v_4$ modulation created by the non-linearity of the suppression with the matter density and size. Nevertheless, for this work we only focus on the $v_2$ coefficient.

We repeat this procedure for every NTC scenario and plot the relative enhancement over the default assumption $\hat{q} \sim \epsilon^{3/4}$ (which corresponds to a reduction of quenching near $T_C$). Note again that only the relative enhancement is meaningful at this stage --- the absolute value of $v_2$ obtained in the computation depends on multiple  factors, among them the pathlength dependence of the jet-medium interaction and the importance of quantum coherence effects, the initial eccentricity distribution of the matter, the amount of viscous entropy production dependent on the value of viscosity over entropy density $\eta/s$ or the precise choice of the decoupling surface for jets from the medium \cite{SysFluid}. By considering the relative enhancement only, many of these uncertainties approximatey drop out.

\begin{figure}
\epsfig{file=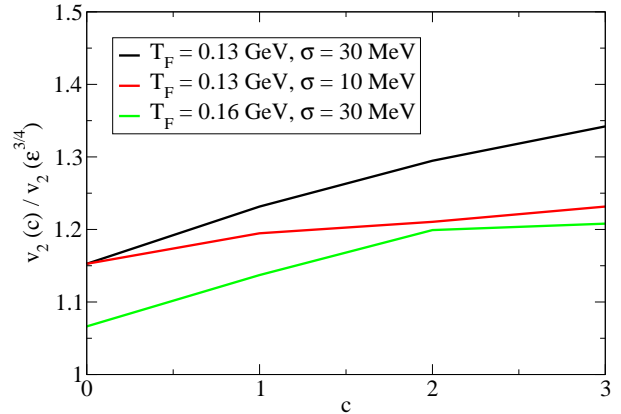, width=8cm}
\caption{\label{F-v2}Relative increase of $v_2$ at $P_T=10$ GeV for 30-40\% central PbPb collisions at 2.76 ATeV for the various scenarios of NTC (see text) relative to the default $\epsilon^{3/4}$ scenario as a function of the NTC strength parameter $c$.}
\end{figure}

The result is shown in Fig.~\ref{F-v2}. A few observations can readily be made: First, for all scenarios tested, the highest enhancement found is 35\%. This is sizable and comparable with e.g. the combined effect of slow thermalization and viscous entropy production \cite{SysFluid}, but smaller than the influence of the spacetime extent of the medium. For weak coupling scenarios \cite{SpreadData,CUJet,YaJEM-D} which tend to underpredict $v_2$ at high $P_T$, NTC is favoured but can not unambiguously be identified as the one dominating factor.

Second, about half of the possible effect already results from not having a reduction of quenching around $T_C$, larger values of $c$ corresponding to more pronounced enhancement still increase $v_2$, but there are indications for a saturation. Third, and perhaps not surprisingly, the effect of NTC is more pronounced the more NTC is probed by the hard parton. Both when the evolution is carried to a lowerdecoupling  temperature  and when the  region in which NTC is effective is increased, a higher relative enhancement of $v_2$ is found.

\section{Discussion}

In this work, the effect of a near $T_C$ enhancement of the parton-medium interaction on both the extraction of the transport coefficient $\hat{q}$ in central collisions and the enhancement of $v_2$ in non-central collisions has been investigated. The results have been obtained in a well-constrained and realistic model combination of fluid dynamics and parton-medium interactions, but given previous systematic investigations across different models \cite{SysEloss,SysFluid} there is reason to expect that relative effect magnitudes are more generally valid.

It was found that the extraction of a value for $\hat{q}(T)/T^3$ from $R_{AA}$ in central collisions in the high temperature region $T > 250$ MeV, i.e. where the quark-gluon plasma (QGP) is expected to exist, is not substantially influenced by any assumptions about the near $T_C$ region. This is fortunate, as it allows to access the physics of the QGP without a full understanding of the phase transition and hadronization. However, other factors, for instance the uncertainty in the total size of the region in which parton and medium interact, pose still a challenge for any precision extraction.

In contrast, $v_2$ as found to be sensitive to NTC as suggested in \cite{Liao,Liao2,Liao1} on a level of a 35\% enhancement. This is in line with the basic findings of \cite{SysFluid} that $v_2$ can generically be expected to increase when energy loss from the leading parton happens later. Comparing weak coupling scenarios with data, such an increase is certainly supported  and indicates that NTC is favoured over a reduction of the interaction near $T_C$. However given the sizable other systematic uncertainties affecting the absolute value of $v_2$, it is difficult to tell whether NTC is required by the data and to unambiguously determine the size of the enhancement. Most likely, an answer to this question will require a systematic picture across several different high $P_T$ observables and will be the topic of a future investigation.

\begin{acknowledgments}
  This work is supported by the Academy researcher program of the
Academy of Finland, Project No. 130472. 
 
\end{acknowledgments}

\end{document}